\documentclass[twocolumn,switch]{article}

\usepackage{arxiv}
\usepackage{mystyle}

\usepackage[utf8]{inputenc} %
\usepackage[T1]{fontenc}    %
\usepackage{hyperref}       %
\usepackage{url}            %
\usepackage{booktabs}       %
\usepackage{amsfonts}       %
\usepackage{nicefrac}       %
\usepackage{microtype}      %
\usepackage{lipsum}         %
\usepackage{graphicx}
\usepackage{natbib}
\usepackage{doi}
\usepackage{multirow}
\usepackage{algorithm}
\usepackage{algorithmic}
\usepackage{amsmath}
\usepackage{cleveref}       %
\usepackage{xcolor}

\definecolor{bluesky}{HTML}{236AAD}
\definecolor{darkblue}{HTML}{034B8D}

\hypersetup{colorlinks=true, linkcolor=purple, urlcolor=darkblue, citecolor=bluesky, anchorcolor=black}

\title{Over 100-fold improvement in the accuracy of relaxed eddy accumulation flux estimates through error diffusion}

\date{August 30, 2024}

\newif\ifuniqueAffiliation
\uniqueAffiliationtrue

\ifuniqueAffiliation %
\author{ \href{https://orcid.org/0000-0002-0141-5262}{\includegraphics[scale=0.06]{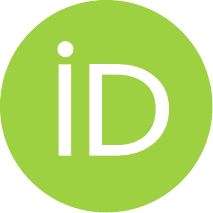}
    \hspace{1mm}Anas Emad}\thanks{} \\
	Bioclimatology\\
University of Göttingen \\ 
Büsgenweg 2, 37077 \\
Göttingen, Germany \\
	\texttt{anas.emad@uni-goettingen.de} \\
}
\fi

\hypersetup{
    pdftitle={Over 100-fold improvement in the accuracy of relaxed eddy accumulation flux estimates through error diffusion},
    pdfsubject={physics.ao-ph},
    pdfauthor={Anas Emad},
    pdfkeywords={Eddy covariance, conditional sampling, slow-response sensors,
    atmospheric exchange, land-atmosphere interactions},
}

\begin{document}
\twocolumn[
\begin{@twocolumnfalse}
\maketitle

\begin{abstract}
Measurements of atmosphere-surface exchange are largely limited by the availability of fast-response gas analyzers; this limitation hampers our understanding of the role of terrestrial ecosystems in atmospheric chemistry and global change. Current micrometeorological methods, compatible with slow-response gas analyzers, are difficult to implement, or rely on empirical parameters that introduce large systematic errors.

Here, we develop a new micrometeorological method, optimized for slow-response gas analyzers, that directly measures exchange rates of different atmospheric constituents, with minimal requirements. The new method requires only the sampling of air at a constant rate and directing it into one of two reservoirs, depending on the direction of the vertical wind velocity. An integral component of the new technique is an error diffusion algorithm that minimizes the biases in the measured fluxes and achieves direct flux estimates.

We demonstrate that the new method provides an unbiased estimate of the flux, with accuracy within 0.1\% of the reference eddy covariance flux, and importantly, allows for significant enhancements in the signal-to-noise ratio of measured scalars without compromising accuracy. Our new method provides a simple and reliable way to address complex environmental questions and offers a promising avenue for advancing our understanding of ecological systems and atmospheric chemistry.
\end{abstract}

\vspace{0.5cm}


\vspace{1cm}
\end{@twocolumnfalse}]

\introduction  %

Surface exchange rates or fluxes of different atmospheric constituents are key
metrics for investigating and understanding the interactions between the
atmosphere and the biosphere \citep{baldocchiHowEddyCovariance2020}. The
uncertainty over the distribution of atmospheric species emissions and losses at
Earth's surface poses a major limitation to our understanding of atmospheric
chemistry and the role of the terrestrial surface in global climate
\citep{arnethTerrestrialBiogeochemicalFeedbacks2010}.
Flux studies provide valuable insights into the cycling of trace elements like
\chem{CO_2}, monitor the emission and deposition of pollutants for public
health and air quality research, and contribute to better representation of
different atmospheric constituents in climate models
\citep{scholesBiosphereatmosphereInteractions2003,fowlerAtmosphericCompositionChange2009a}.

Micrometeorological methods are employed to measure atmospheric exchange in the
lower atmosphere near the surface, where the transport is predominantly turbulent,
causing gases to diffuse rapidly from and to the surface. The flux near the
surface is determined as the average of the product of the vertical wind
velocity $w$ and the scalar
concentration $c$, $\overline{w\,c}$, where the overline denotes the average over
a period of time (typically 30 minutes to 1 hour). The direct calculation of
the flux $\overline{w\,c}$ using the eddy covariance method, is widely recognized as the most
direct and reliable way for measuring atmospheric fluxes at the scale of plant
canopies \citep{dabberdtAtmosphereSurfaceExchangeMeasurements1993,
hicksMeasurementFluxesLand2020}.

Atmospheric exchange occurs over various spatial and temporal scales,
requiring high-frequency measurements (above 10 Hz) of wind velocity and
scalar concentration to capture all flux-carrying motions. 
This high sampling requirement has limited the eddy covariance method to few
atmospheric constituents, like water vapor and carbon dioxide, which have
fast-response gas analyzers. Yet, many constituents
critical to atmospheric chemistry and ecosystem research, such as stable
isotopes, nitrogen oxides, and volatile organic compounds, only
have slow-response analyzers. For these, several alternative
micrometeorological methods compatible with slower analyzers
have been developed \citep{rinneAlternativeTurbulentTrace2021}. 
Among these alternatives, the true eddy accumulation (TEA) method offers a
direct and accurate way to measure the flux $\overline{w\,c}$ without requiring
fast-response analyzers. TEA achieves this by sampling air with a mass
proportional to the vertical wind velocity and then directing it into one of two
reservoirs based on the vertical wind direction. Therefore, the product $w\times
c$ is realized physically by proportional sampling and the mean is achieved by
sample accumulation. The flux is then calculated from the accumulated mass
difference between the two
reservoirs
\citep{desjardinsDescriptionEvaluationSensible1977,hicksSimulationEddyAccumulation1984}.
However, the accurate and fast control of air mass flow rate at the necessary dynamic range
required for TEA implementation presents a significant challenge. This has
limited the number of successful implementations of the TEA method and has
prevented its widespread adoption \citep{siebickeTrueEddyAccumulation2019}.
A simplification of the eddy accumulation method that does not require
proportional mass flow control was proposed by
\cite{busingerFluxMeasurementConditional1990}, which draws parallels to flux
gradient methods by linking the flux with the difference between mean updraft and
downdraft scalar concentrations.
Unlike TEA, the relaxed eddy accumulation method (REA)
does not require proportional mass flow control; instead, air samples are
collected at a constant flow rate into updraft and downdraft reservoirs
depending on vertical wind direction. The flux in REA is found as the product of
the difference between mean accumulated scalar concentrations ($\Delta C$), wind standard
deviation $(\sigma_w)$, and an empirical coefficient $\beta$.
The interpretation and determination of the empirical coefficient $\beta$ have
remained the primary questions concerning the REA method
\citep{bakerConditionalSamplingRevisited2000,
yukioStatisticalDerivationFundamental2004,
fotiadiMethodologicalDevelopmentConditional2005,
ruppertScalarSimilarityRelaxed2006,
katulEjectiveSweepingMotions2018,
voglChoosingOptimalFactor2021}.
While the theoretical value of $\beta$ derived  
assuming a Gaussian distribution for wind and scalar
is $\beta \approx 0.63$ \citep{wyngaardParameterizingTurbulentDiffusion1992},
observed average values of $\beta$ typically range between 0.47 and 0.63 and
exhibit substantial run-to-run variability
\citep{gaoVerticalChangeCoefficient1995,katulInvestigationConditionalSampling1996,tsaiEvaluationRelaxedEddy2012,
grelleAffordableRelaxedEddy2021}.
These variations are attributed 
to diverse stability conditions, scalars, and site-specific factors
\citep{sakabeEmpiricalCoefficientRelaxed2014,ammannStabilityDependenceRelaxed2002}. 
The absence of a reliable method for estimating or parametrizing $\beta$ has
made it a major source of uncertainty in REA flux measurements, potentially
introducing biases of up to 20\% of the flux
\citep{oncleyVerificationFluxMeasurement1993,gaoVerticalChangeCoefficient1995}.

Despite the high measurement uncertainty associated with the variability of the
coefficient $\beta$,
the REA method remains widely used for atmospheric exchange measurements due to
its simple implementation requirements and its ability to increase scalar
signal-to-noise ratios by excluding lower wind speeds using a
deadband. REA has been applied to measure the fluxes of a wide array of
atmospheric constituents.
These include stable isotopes,
hydrocarbons,
methane,
volatile organic compounds, 
ammonia, sulfate,
nitrous oxide, 
mercury,
aerosol numbers and dry deposition,
halocarbon fluxes,
peroxyacetyl nitrate,
sodium chloride particle,
and benthic solute fluxes in aquatic environments
\citep{
beverlandMeasurementMethaneCarbon1996,
haapanalaMeasurementsHydrocarbonEmissions2006,
bowlingDynamicsIsotopicExchange1999,
zhuAircraftbasedVolatileOrganic1999,
patteyMeasurementIsopreneEmissions1999,
ciccioliRelaxedEddyAccumulation2003,
matsudaDryDepositionPM22015,
hensenIntercomparisonAmmoniaFluxes2009,
grelleAffordableRelaxedEddy2021,
skovFluxesReactiveGaseous2006,
meyersFluxesAmmoniaSulfate2006,
osterwalderDualinletSingleDetector2016,
heldRelaxedEddyAccumulation2008,
gronholmMeasurementsAerosolParticle2007,
hornsbyRelaxedEddyAccumulation2009,
Ren2011,
moravekApplicationGCECDMeasurements2014,
meskhidzeContinuousFlowHygroscopicityresolved2018,
calabro-souzaNewTechniqueResolving2023,
riedererPrerequisitesApplicationHyperbolic2014}.

In this paper, we propose a new direct eddy accumulation method that combines
the simple requirements of relaxed eddy accumulation with the accuracy and
robustness of the conventional eddy covariance technique. The new method employs
error diffusion to achieve reliable and direct flux estimates with minimal bias.
We begin with a novel theoretical derivation of conditional sampling,
formulating sample accumulation as a quantization problem and demonstrating that
relaxed eddy accumulation is a special case of this formulation. We identify an
unaccounted portion of the flux in the REA method arising from the correlation
of the quantization error with the scalar concentration. We then introduce an
error diffusion algorithm to randomize this quantization error, leading to
unbiased flux estimates.
This is followed by a detailed examination of the signal and noise shaping
properties of error diffusion and a discussion of how it effectively minimizes
flux errors associated with quantization errors.
Finally, we test the performance of our method through numerical simulations. We
explore the ideal quantization parameters and discuss the implementation details
of this new approach.

\section{Theory}

\subsection{Derivation of quantized eddy accumulation}

In the following, we derive a general equation for conditional sampling flux measurement techniques and show how relaxed eddy accumulation (REA) can be formulated as a special case of this equation.

We consider the flux $F$ (\unit{mol~m^{-2}~s^{-1}}) of an atmospheric
constituent $c$, such as \chem{CO_2}, as 

\begin{equation}
    \label{eq:general-flux}
    F = \overline{wc} = \overline{w'c'} + \bar{w}\bar{c}
\end{equation}

Here, $w$ represents the vertical wind velocity ($\mathrm{m}\, \mathrm{s}^{-1}$), and $c$ represents the molar density ($\mathrm{mol}\, \mathrm{m}^{-3}$) of the scalar in question. Overlines indicate averaging that follows Reynolds averaging rules, while the primes indicate deviations from the mean. The previous equation can be derived from conservation laws. For a detailed discussion of surface flux equations, refer to
\cite{finniganReEvaluationLongTermFlux2003} and
\cite{fokenEddyCovarianceMethod2012}. 
The definition in Eq.~\ref{eq:general-flux} is chosen as eddy accumulation
methods measure, by definition, the term $\overline{wc}$.  However, similar to
eddy covariance (EC) measurements, the term $\bar{w}\bar{c}$ is typically biased
due to inaccuracies in $w$ \citep{emadTrueEddyAccumulation2023}.  
The estimation of the physical $\bar{w}$ associated with non-turbulent transport
is still relevant, as in EC measurements, and depends on the used concentration
index \citep{kowalskiBoundaryConditionVertical2017,kowalskiDisentanglingTurbulentGas2021}.
However, these aspects do not affect our subsequent analysis.

We apply a quantizer function $Q(w)$ to restrict the possible values of $w$ to a
finite, discrete set. To ensure high flexibility and ease of implementation in
flux measurements, we consider a non-uniform quantizer with three levels:
$(-w_{f}, \,0,\, +w_{f})$. These quantization levels do not need to be uniformly
spaced. The quantizer function compares measured wind speeds with a quantization
threshold, denoted as $q_t$. If the measured magnitude of wind speed exceeds
this threshold, the quantized value of $w$ becomes the full-scale value $w_{f}$
or $-w_{f}$ depending on the direction of $w$. The quantizer function is
expressed as

\begin{equation}
    \label{eq:quantizer-function}
    Q(w, ~q_t,~ w_{f}) = 
    \begin{cases}
        \mathrm{sign}(w)\, w_{f}, & \text{if}\, |w| > q_t  \\
        0, & \text{if}\, |w|\leq q_t 
    \end{cases}
\end{equation}

We define the quantization error $\varepsilon$ as the difference between the
true value of $w$ and the quantized output $w_q$ such that

\begin{equation}
    \label{eq:quantization-error-definition}
    {\varepsilon} = w - w_q.
\end{equation}

Consequently, the flux can be written as the sum of the flux of the quantized wind and the flux of the quantization error

\begin{equation}
    \label{eq:flux-quantized}
    \overline{w\,c}  = \overline{c (w_q + \varepsilon )} =  
    \overline{w_q\, c} +
    \overline{\varepsilon\, c}.
\end{equation}

We are more interested in the term $\overline{w'\,c'}$, we rearrange the
previous equation and decompose $w_q$ and $\varepsilon$ into mean and fluctuating components. We obtain

\begin{equation}
    \label{eq:flux-quantized-fluctuations}
  \overline{w' c'} = \overline{w_q\, c} + \overline{\varepsilon'\, c'} -
  \overline{w_q}~\overline{c}
\end{equation}

We express the flux of quantized wind, $\overline{w_q\, c}$, using the law of total expectation as the expectation of the conditional mean of the random variable $c \times w_q$ partitioned on the variable $I$.  The variable $I$ divides the probability space of $w_q\,c$ into distinct, non-overlapping partitions $\{I_1, I_2, \ldots, I_k\}$. The expectation of $\overline{c\,w_q | I}$ is calculated as a sum over these partitions, each weighted by the probability of its occurrence. Thus, Equation \(\ref{eq:flux-quantized-fluctuations}\) is written as

\begin{equation} 
    \label{eq:general-qea-definition}
    \overline{w'c'} =
    \sum_{j=1}^{k} \overline{c_{I_j} w_{q{I_j}}}
    \times P(I_j) + 
    \overline{\varepsilon'\, c'} - 
    \overline{w_q}~\overline{c} 
\end{equation}

The usefulness of this formulation becomes evident when we consider that for a
any quantization level, $\overline{w_q\,c} = \overline{w_q}~\bar{c}$. This
allows us to obtain the covariance term $\overline{w_q\,c}$ directly from
measurements of the mean $\overline{c_{I_j}}$ in different partitions instead of
requiring the high-frequency measurements of $c$. Additionally, the
realization of $\overline{w_{qI_j}\,c_{I_j}}$ for a given $I_j$ by air
accumulation is more straightforward as it requires no proportional control of
the airflow.

If we consider a simple partitioning scheme based on the direction of vertical
wind velocity, $I = \mathrm{sign}(w)$, and choose $\sigma_w$, the standard
deviation of $w$, as the full-scale value for our quantizer, we obtain the
quantized time series $w_q$ with the same length as $w$ but with values that can
be either $-\sigma_w$ or $\sigma_w$. The flux can then be obtained from
Eq.~\Ref{eq:general-qea-definition} as

\begin{equation}
    \label{eq:REA-flux-exact}
    \overline{w'\,c'} = 
    \sigma_w \, \overline{c^\uparrow}   \frac{n^{\uparrow}}{n} - 
    \sigma_w \,\overline{c^\downarrow} \frac{n^{\downarrow}}{n} -
    \bar{c} \,\sigma_w \,\frac{n^{\uparrow} - n^{\downarrow}}{n} +
    \overline{c' \, \varepsilon'}
\end{equation}

We notice that when $n^\uparrow = n^\downarrow$, the previous equation is reduced to the familiar formula

\begin{equation} 
    \label{eq:REA-AS-QEA-flux}
    \overline{w'\,c'} = 
    \frac{1}{2}\, \sigma_w \,\Delta C + \overline{c'\, \varepsilon'} 
\end{equation}

Comparing this simplified formula to the common equation used to define REA flux
reveals that the measured quantities in REA correspond directly to the flux
resulting from quantized wind speed, specifically $\sigma_w\,\Delta C
=2\,\overline{w_q'\, c'}$.  This comparison highlights that REA ignores the flux
associated with the quantization error, $\overline{c'\, \varepsilon'}$, and
estimates the true flux from the quantized flux through the empirical parameter
$\beta$ which is effectively the ratio of the true flux to the quantized flux
$\beta = {\overline{w'\,c'}}/{(2\,\overline{w'_q\,c'})}$.

Equation \ref{eq:REA-flux-exact} highlights two fundamental limitations in
conventional REA methods. First, they neglect the flux component
$\overline{\varepsilon'\,c'}$, and instead estimate its value empirically
through the parameter $\beta$. Second, the quantization does not preserve the
mean wind, $\overline{w_q} \neq \overline{w}$, as this would require $n^\uparrow
= n^\downarrow$ which is rarely the case. The nonzero $\overline{w_q}$ leads to
a biased flux estimate due to the inability to account for the term
$\overline{c}~\overline{w_q}$ when using a quantization threshold different
from zero (a deadband).

This raises the question of whether using more optimal quantization —by
selecting better values for the quantization threshold and full-scale value—
could lead to improved estimates of the term $\overline{\varepsilon'\,c'}$.
Essentially, the aim of varying $\beta$ parameterization schemes and adjusting
deadbands in REA is to address this specific issue. However, these modifications
had achieved limited success.

Rather than improving the estimates of the term $\overline{\varepsilon'\,c'}$,
we propose to reduce its value to zero by randomizing the quantization error to
eliminate its correlation with the scalar concentration. The randomization can
be achieved by adding pseudorandom noise to the wind signal. The added noise is
formed from past errors through a feedback loop, a process known as error
diffusion. A simple form of error diffusion can be expressed as

\begin{equation}
    \label{eq:simple-error-diffusion}
    w_q[n] = Q\left(w[n] - {\varepsilon}[n-1]\right),
\end{equation}

where $w_q[n]$ is the quantized wind speed at time $n$, $w[n]$ is the measured
wind speed, and $\varepsilon[n-1]$ is the quantization error at the previous
time step.  This approach effectively eliminates the correlation between 
the quantization error and the scalar concentration. 
Therefore, making the measured flux of quantized wind speed,
$\overline{w_q'\,c'}$, equal to the true flux, which allows to obtain direct
flux estimates and eliminates the need for the parameter $\beta$.

The addition of noise to improve the quality of quantized signals is a well-known technique in signal processing. Error diffusion is commonly employed in image and audio processing applications to enhance the perceptual quality of quantized signals \citep{knoxEvolutionErrorDiffusion1999,kiteDigitalHalftoning2D1997, escbbachThresholdModulationStability2003}.  We refer to this variant of eddy accumulation as quantized eddy accumulation (QEA) with error diffusion.

\subsection{Analysis of error diffusion}
We are interested in analyzing the signal and noise-shaping behavior of error
diffusion. 
We consider quantization and error diffusion as a system that takes an input signal
$x(n)$ and produces an output signal $y(n)$, where the input signal is the
vertical wind velocity and the output signal is the quantized wind velocity.
We first show that error diffusion acts as a high-pass filter on the
noise, then we discuss the optimal choice of the error diffusion filter.

The first step of quantization and error diffusion is forming the modified
input. This is achieved by subtracting "diffused" past error terms from the
input. We write a
similar, but more general form of Eq.~\ref{eq:simple-error-diffusion} as

\begin{equation}
    \label{eq:modified-input-time-domain}
    u(n) = x(n) - \sum_{k \in \mathcal{O}} h(k)\,e(n - k).
\end{equation}
Here, $u(n)$ represents the modified input, $x(n)$ corresponds to the input of
the system such as vertical wind velocity. 
The diffusion of past errors $e(n-k)$ is represented by the linear weighting
filter, denoted as $h$. $\mathcal{O}$ is the causal
support set that does not include 0, signifying that it only includes past
errors. 

In the second step, the output of the system $y(n)$ is calculated by applying the quantization function $Q(.)$ to the modified input.
\begin{equation}
    y(n) = Q(u(n)),
\end{equation}
In our use case, the output of the system will be the quantized vertical wind
velocity that had past errors integrated into it.
The quantizer error $e(n)$ is defined as the difference between the output, $y(n)$, and the modified input
\begin{equation}
    \label{eq:quantizer-error-time-domain}
    e(n) = y(n) - u(n),
\end{equation}
note that this is different from the quantization error defined earlier.
We substitute Eq.~\ref{eq:modified-input-time-domain} into Eq.~\ref{eq:quantizer-error-time-domain}, we find

\begin{equation}
    \label{eq:error-complete-time-domain}
    e(n) = y(n) - x(n) + \sum_{k \in \mathcal{O}} h(k) e(n - k)
\end{equation}

The previous equation highlights the recursive character of error diffusion, where the error at any given moment in time depends on the entire history.
To analyze this relation in the frequency domain, we apply the $z$-transform
which converts discrete-time signals from the time domain into the $z$-domain
\citep{smithIntroductionDigitalFilters2007}. After applying the $z$-transform
and rearranging, we can write Eq.~\ref{eq:error-complete-time-domain} as

\begin{equation}
    \label{eq:quantization-error-z-domain}
    Y(z) - X(z)  = (1 - H(z)) E(z).
\end{equation}
Here, $H(z) = \sum h(k) z^{-k}$, is the filter applied to past errors in the
frequency domain. 
The variable $z$ is a complex exponential reflecting the frequency components of
the signal and is defined as
$z=\exp({j\omega T})$,
where $\omega$ is the angular frequency in radians per sample, and $T$ is the
sampling interval.
Considering that $G(z) = Y(z) - X(z)$ represents the total
quantization error, which corresponds to the output of the noise transfer
function, we find the transfer function describing the frequency shaping of the
noise by rearranging the previous equation to be

\begin{equation}
    \label{eq:quantization-error-transfer-function}
    \mathrm{NTF}(z)= \frac{G(z)}{E(z)}  = 1 - H(z) 
\end{equation}

Equation~\ref{eq:quantization-error-z-domain} indicates that the total
quantization error $\varepsilon$, defined as the difference between the system
output and its input, consists of two errors: one is a filtered version of the
other.
The first error component is the quantizer error $E(z),$ which we defined as the
difference between the output and the modified input.  The second error
component is the filtered quantizer error, which we call the diffusion error
$E(z)H(z)$.

Error diffusion can be viewed as an error minimization process. It aims to
reduce the difference between the input and the output by adjusting the weights
of a linear filter \citep{escbbachThresholdModulationStability2003}. In our
specific application, our primary interest lies not in minimizing the quantization error
itself, but rather in minimizing its correlation with the scalar concentration $c$,
therefore, we want to choose $H(z)$ such that the filtered output $H(z)E(z)$ is
as close as possible to $E(z)$. 
Given that $H(z)$ has to be causal, we find the unit delay filter, denoted as
$H(z) = z^{-1}$, as the optimal solution for minimizing the correlation of the
quantization error with the scalar.
The unit delay filter allows all frequencies to pass unattenuated
and introduces a linear phase shift, which, in this case, results in a
one-sample delay of the input.  This means that the error from the previous time
step is added in its entirety. 
The rationale for this choice
is that $e(n)$ is not completely random; instead it has an autocorrelation
structure due to the recursive nature of error diffusion. Therefore, the closest
version of $e(n)$ that can be achieved using a causal filter is a
one-sample-delayed version of itself.  %
The previous equation with $H(z)$ as the unit delay filter
indicates that the mean of the output equals the mean of the input
because the filter coefficients sum to one; therefore, at zero frequency, $H(1)
= 1$.

To investigate the behavior of flux errors, we can express the quantization
error covariance term in the time domain as:

\begin{equation}
    \overline{\varepsilon'\, c'} =  \overline{e' c'} - \overline{(h*e)' c'},
\end{equation}
where $*$ denotes convolution.  
Given that diffusion error represents a
one-sample delayed version of $e(n)$ and based on the definition of the
cross-covariance function between $e$ and $c$, denoted as $R_{ec}(n)$, we can
express the error covariance as

\begin{equation}
    \label{eq:error-by-cov-function}
    \overline{\varepsilon'\, c'} = R_{ec}(0) - R_{ec}(1)
\end{equation}

Typically, $e$ is assumed to be white noise for a uniform quantizer
\citep{kiteModelingQualityAssessment2000}.
However, for a non-uniform quantizer, $e$ exhibits 
correlation with $w$ and $c$.
To address this issue, we
can apply error diffusion per wind direction, effectively treating the updraft
and downdraft as separate signals. 
This approach enables non-uniform quantization without introducing correlation,
which offers more flexibility in choosing quantization thresholds and full-scale
values.
Modeling the quantizer directly is challenging due to its nonlinearity, which
complicates the characterization of the cross-correlation function $R_{ec}$,
especially when applying diffusion per wind direction. However, $e$ is
expected to exhibit some autocorrelation structure due to the recursive nature
of error diffusion. This autocorrelation further contributes to minimizing the
flux error in Eq.~\ref{eq:error-by-cov-function}. The experimental simulation
results in the results section confirm this analysis.

\subsection{Calculation of QEA fluxes}
We show here how the flux is calculated in QEA for the simple case of
partitioning based on the sign of $w$. In this case, the quantized flux in
equation~\ref{eq:general-qea-definition} becomes

\begin{equation}
    \overline{w_q\,c} =
    \overline{c^\uparrow w_q^\uparrow} ~p^\uparrow + 
    \overline{c^\downarrow w_q^\downarrow}  ~p^\downarrow,
\end{equation}
where the arrows indicate updraft and downdraft, and $p$ is the probability of wind direction being in the given direction.  After the end of the averaging interval, typically spanning 30 minutes to 1 hour, the flux $\overline{w'c'}$ is obtained from Eq.~\ref{eq:general-qea-definition} as

\begin{equation}
    \label{eq:flux-calculation-general}
    F_{\text{total}} = \overline{w'c'} = 
    \overline{c^{\uparrow}}~\overline{w_q^\uparrow} \frac{n_q^{\uparrow} }{n}
    +
    \overline{c^{\downarrow}}~\overline{w_q^\downarrow} \frac{n_q^{\downarrow}}{n}
    - \bar{c}~\overline{w_q}
    + 
    \overline{\varepsilon'\, c'},
\end{equation}

where $\overline{c^{\uparrow}}$ and $\overline{c^{\downarrow}}$ represent the
average scalar concentrations in the updraft and downdraft reservoirs, while
${n_q^{\uparrow}}$ and ${n_q^{\downarrow}}$ indicate the counts of occurrences
with the quantized wind direction as updraft and downdraft, respectively. The
variable $n$ represents the total number of vertical wind velocity samples
within the averaging interval.

Equation \ref{eq:flux-calculation-general} includes the term
$\bar{c}~\overline{w_q}$, which requires knowledge of the average scalar
concentration $\overline{c}$ within the averaging interval.  Estimating this
term is a common challenge in eddy accumulation methods and arises due to the
presence of a biased non-vanishing mean vertical wind velocity.  The root of
this issue is that samples are accumulated in real time without full knowledge
of wind statistics throughout the averaging interval, making it impossible to
ensure that $\bar{w}$ is zero.  Solutions to this problem typically involve
estimating $\bar{c}$ from available measurements and the properties of atmospheric
transport \citep{emadTrueEddyAccumulation2023}.  We note that for QEA with a
zero quantization threshold, $\overline{c}$ can be calculated directly from the
measurements of $\overline{c^\uparrow}$ and $\overline{c^\downarrow}$.
To accommodate the estimation of $\overline{c}$ from available measurements, we
express the total flux, $F_{\text{total}}$ as the product of the base flux,
$F_{\text{1}}$, a correction factor, $A_{\bar{w}}$, to account for non-zero mean
wind velocity, and the quantization error covariance component
$F_{\text{error}}$

\begin{equation}
    \label{eq:flux-calculation-total}
    F_{\text{total}} = 
    F_{\text{1}} \cdot A_{\bar{w}} + F_{\text{error}}
\end{equation}

The base flux, $F_{\text{1}}$, is derived from measurable quantities

\begin{equation}
    \label{eq:base-flux-f1}
 F_{\text{1}} = \overline{c^{\uparrow}}~\overline{w_q^\uparrow}
    \frac{n_q^{\uparrow}}{n} +
    \overline{c^{\downarrow}}~\overline{w_q^\downarrow}
    \frac{n_q^{\downarrow}}{n} -
    \left(\overline{c^{\uparrow}}~\frac{n_q^{\uparrow}}{n_q} +
    \overline{c^{\downarrow}}~\frac{n_q^{\downarrow}}{n_q}\right)~\overline{w_q},
\end{equation}
where $n_q$ is the count of occurrences where $w_q \neq 0$ and $n$ is the total sample count in the averaging interval.

The correction factor, \( A_{\bar{w}} \), adjusts for the discrepancy between
the measured average concentration, \( \bar{c} \), and its true value, with
higher wind variability leading to reduced error, and can be estimated as

\begin{equation}
    \label{eq:A-correction-factor}
 A_{\bar{w}} = \frac{1}{1 - \alpha_{\text{c}}
\frac{\overline{w_q}}{\overline{|w_q|}}},
\end{equation}
where $\alpha_c$ is the atmospheric transport asymmetry coefficient defined as the ratio ${\overline{c'\,|w_q'|}}/{\,\overline{c'\,w_q'}}$ and accounts for the correlation when estimating $\overline{c}$ from $\overline{c\,|w_q|}$. The value of $\alpha_c$ can be estimated analytically or empirically \citep{emadTrueEddyAccumulation2023}.  Lastly, \( F_{\text{error}} \) represents the error in the flux due to the quantization error previously discussed, typically less than 0.1\% of the flux.  If the units of measured $\overline{c^\uparrow}$ and $\overline{c^\downarrow}$ are dry mole fractions \unit{mole \cdot mole^{-1}}, we divide the resulting flux by the molar volume of dry air $V_{md}$ (\unit{m^3\cdot mole^{-1}}) to obtain the flux density in units of \unit{mol \cdot m^{-2} \cdot s^{-1}}.
An implementation and examples of flux calculation for this method are provided in the supplementary material.

\section{Methods}
\subsection{Simulation data and parameters}
We used a high-frequency dataset of eddy covariance (EC) measurements collected
from two contrasting ecosystems as input for our simulation. The first is an
ideal flat agricultural field at the Thünen Institute in Braunschweig, Germany
(52.30° N, 10.45° E), where wind measurements were obtained using a sonic
anemometer (uSonic-3 Class A, Metek GmbH, Elmshorn, Germany) and an open-path
infrared gas analyzer (LI-7500A, LI-COR Biosciences Lincoln, NE, USA) at 10 Hz
sampling frequency. Data from this site covered two periods in 2020, from June 1
to July 15 and from October 1 to November 10, resulting in 3,146 30-minute
averaging intervals across 65 days.
The second site is an old-growth forest in the Hainich National Park (ICOS
DE-Hai) in Thuringia, Germany (51.08° N, 10.45° E). Here, measurements were
taken using a 3D sonic anemometer (Gill-HS, Gill Instruments Limited, Hampshire,
UK) and an enclosed-path infrared gas analyzer (LI-7200, LI-COR Biosciences
Lincoln, NE, USA). Data used at the Hainich site covered two periods in 2022,
from June 1 to June 31 and from August 11 to September 15, with a total of 3,490
30-minute averaging intervals over 72 days.
Further details on the sites and instrumentation specifications can be found in
\citep{emadTrueEddyAccumulation2023a} and \citep{Knohl2003} respectively.

Using this dataset, we simulated flux measurements for three variables:
\chem{CO_2}, \chem{H_2O}, and air temperature measured with a sonic anemometer
($\theta$).  Our simulation included the newly developed quantized eddy
accumulation and six conventional relaxed eddy accumulation variants, serving as
baselines for performance benchmarking under the same constraints.  Fluxes for
the different methods were calculated based on 30-minute averaging intervals, and
basic quality controls were applied to the calculated fluxes.  Quality control
measures included flagging periods where stationarity tests failed or where the
friction velocity was below 0.1 \unit{m\,s^{-1}}
\citep{fokenPostFieldDataQuality2005}.

The simulation aimed to assess the performance of the QEA method with error
diffusion and determine its optimal parameters for various key objectives. 
The performance evaluation aimed to quantify flux errors associated with
these methods, relative to the reference eddy covariance method. This assessment
included both systematic and random components of error. Here, we consider
systematic errors to be those correlated with the measured flux value.  Flux
errors are defined as the difference between the measured flux using a given
method and the reference eddy covariance flux.  To provide a relative measure of
the systematic error that works across different scalars, we chose the slope of
a linear regression of the flux error against the flux value as a metric to
quantify systematic errors. 

Random errors, in contrast, are not correlated with the magnitude of the flux,
although their variance may still show a correlation with the flux value.  The
impact of random errors is generally less severe since they tend to diminish
with averaging.
The influence of random errors can be estimated by calculating
the standard deviation of the errors around the measured flux value, providing a
measure of the method’s uncertainty. 
However, as we aim to establish general relative measures for the uncertainty of
the compared methods rather than specifying values for a given scalar, We define
the normalized error $E_{\mathrm{norm}}$ as the ratio of the flux error to the
product of the standard deviations of the vertical wind velocity and the scalar 
\begin{equation}
    \label{eq:normalized-error}
    \mathrm{E}_{norm} = \frac{F_{\text{meas}} -
    F_{\text{ref}}}{\sigma_w \,\sigma_c},
\end{equation}
where $F_{\text{meas}}$ is the measured flux for a certain method, $F_{\text{ref}}$ is the reference eddy covariance flux.
This definition essentially transfers the error to be expressed as error in the correlation coefficient $\rho_{wc}$. 
The standard deviation of the normalized error can be used as a non-dimensional
measure of the uncertainty of the method
\begin{equation}
    \label{eq:nondim-uncertainty}
    u_{\text{nd}} = \sqrt{\mathrm{Var}(E_{\text{norm}})}
\end{equation}
The usefulness of this expression is that it facilitates comparisons of
uncertainty across different scalars as the correlation coefficient is seen as
an independent measure of atmospheric transport.  Additionally, it enables the
calculation of the relative uncertainty of the flux when divided by the
correlation coefficient $\rho_{wc}$. For example, a non-dimensional uncertainty
of 0.01 implies that, if $\rho_{wc} = 0.4$, there is a 95\% probability that the
flux will be within $\pm 5\%$ of the measured value, assuming a normal
distribution of the error.

While we use EC fluxes as our reference, it is important to note that
EC fluxes have limitations, particularly during low turbulence or over complex
terrain. Eddy accumulation methods are designed to address issues of sensor
response time and signal strength, aiming to match EC flux measurements when
using slow-response sensors. However, these methods cannot solve fundamental
problems unrelated to sensor limitations, such as violations of EC assumptions.
We therefore use EC as a reference to evaluate the performance of the eddy
accumulation methods.

Choosing a method for estimating or parametrizing $\beta$ poses a common
challenge for REA methods, as there are no clear recommendations that
universally apply across all sites and atmospheric conditions. We chose six
methods to estimate $\beta$ as recommended by recent work on REA
\citep{voglChoosingOptimalFactor2021}.  These treatments included the commonly
used $\beta$ calculated from sonic temperature
$\beta_{ts}=\overline{w'\theta'}/(\Delta \theta \, \sigma_w)$, a constant
$\beta$ estimated as the median of $\beta$ calculated from air temperature
$\beta_{ts-median}$, and $\beta$ calculated from vertical wind velocity
statistics as $\beta_{w} =\sigma_w/\Delta w$ following
\cite{bakerConditionalSamplingRevisited2000}.  
Additionally, we used for each of the previous methods two dynamic linear
deadband settings, namely, $0.5\,\sigma_w$ and $0.9\,\sigma_w$.
The $\beta$ values calculated using these methods were subsequently used to
compute REA fluxes through the equation $F_{\text{REA}} = \beta \, \sigma_w \,
\Delta C$. Each method is identified by combining the $\beta$ estimation
technique with the chosen deadband setting, such as
$\mathrm{REA}~(\mathrm{db}~0.5\sigma)~\beta_{\mathrm{ts-median}}$.

Quantized eddy accumulation with error diffusion was simulated with a wide range
of parameters.  The algorithm used to simulate QEA with error diffusion is
detailed in Section~\ref{sec:implementation}.  For performance comparison with
REA variants, we selected the parameter pair ($q_t=2.4 \sigma_w,~w_{f}=4
\sigma_w$). 
The full code of the simulation and the analysis are available at
\cite{emadDataSupportCode2024}.

\section{Results and Discussion}\label{sec2}

\subsection{Quantized eddy accumulation with error diffusion}

We have developed a general framework for conditional sampling methods in
Eq.~\ref{eq:general-qea-definition} by combining quantization of vertical wind
speed with conditional sampling. Conventional relaxed eddy accumulation method
was shown to be a special case of this framework when setting the conditioning
variable $I$ to $\mathrm{sign}(w)$ and quantizing wind speed into two discrete
levels that correspond to the standard deviation of the wind $-\sigma_w$ and
$+\sigma_w$ as seen in Eq.~\ref{eq:REA-flux-exact}.
The parameter $\beta$ is found to be fundamentally linked to an unaccounted flux
term that is carried with quantization errors arising from non-optimal
quantization. This connection explains the difficulty in estimating $\beta$ as
it involves estimating a covariance term based on the conditional mean of scalar
concentration and wind statistics.

We proposed to use error diffusion to eliminate the covariance term
$\overline{\varepsilon'\,c'}$ by adding pseudorandom noise to $w$ in a feedback
loop. Eliminating this term makes the quantized flux $\overline{w_q\,c}$ equal
to the true flux $\overline{w'\,c'}$ and eliminates the need for the empirical
coefficient $\beta$.  
The error diffusion algorithm is seen as an error minimization mechanism, as
pseudorandom noise is introduced into the signal to reduce the error correlation
with the scalar.  We found that a simple variant of error diffusion where the
complete quantization error is added to the next sample is optimal for
minimizing the correlation between the quantization error and the scalar
concentration.  This corresponds to using the unit delay filter $H(z) = z^{-1}$
in Eq.~\ref{eq:quantization-error-z-domain}.

QEA with error diffusion offers the advantage of increasing the difference
between mean accumulated concentrations in updraft and downdraft reservoirs
$\Delta C$, thereby improving the signal-to-noise ratio.
The quantization of wind with a threshold larger than zero, similar to employing
a deadband in REA, serves the purpose of filtering out low wind speeds that are
associated with smaller scalar fluctuations.  
This concentration enhancement is especially useful when
measuring atmospheric constituents with low flux. Importantly, this
concentration enhancement does not introduce additional uncertainty to the flux
as error diffusion performance is largely independent of the quantization
threshold as is shown in Fig.~\Ref{fig:parameter-space-heatmap}.

\begin{figure}[t]
\centering
\includegraphics[width=8.2cm]{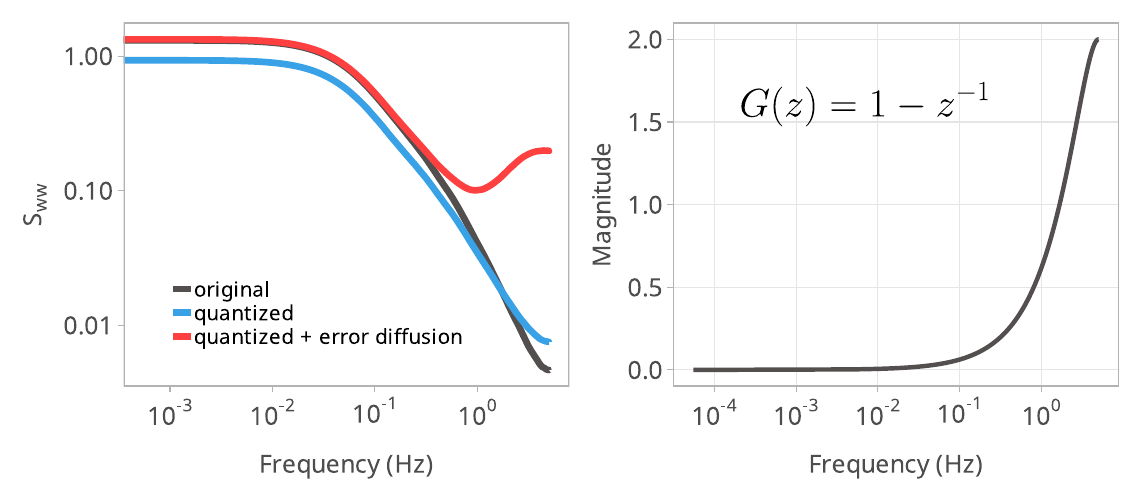}
\caption{
Spectral analysis of error diffusion. (a) Normalized averaged power spectral
density of vertical wind velocity for a typical 30-minute interval.  The
'original' curve represents the original spectrum, while the 'quantized' and
'quantized + error diffusion' curves illustrate the effects of quantization and
subsequent error diffusion on the signal. The noise shaping behavior is
demonstrated by the deviation of 'quantized + error diffusion' spectrum from the
'original', particularly in the higher frequency range. (b) Magnitude response
of the noise transfer function, displaying the high-pass characteristics of the
$G(z) = 1 - {z^{-1}}$ filter, with a steady increase in magnitude with
frequency, demonstrating the noise shaping capability.  The displayed spectra
are averaged from 50 samples taken from random 30-minute intervals that meet the
quality standard $|\rho_{wc}| > 0.3~\mathrm{and}~\sigma_w > 0.4$.
}
\label{fig:noise-ntf}
\end{figure}

The analysis of error diffusion in the $z-$domain reveals that the quantizer
noise is pushed towards the higher end of the spectrum, as shown in
Fig.~\Ref{fig:noise-ntf}. Consequently, the spectrum of the quantization error
exhibits characteristics resembling blue noise, with a significant concentration
of power in the high-frequency range.
When we consider the unit delay filter $1/z$ for the error diffusion filter.
The noise transfer function in Eq.~\ref{eq:quantization-error-transfer-function}
becomes $\mathrm{NTF}(z) = 1 - {z^{-1}}$. $\mathrm{NTF}(z)$ exhibits a frequency
response that amplifies higher frequencies, as indicated by the magnitude plot
in Fig.~\ref{fig:noise-ntf}. The gain increases linearly with frequency, peaking
at twice the input signal’s amplitude at the Nyquist frequency. This
characteristic introduces additional energy into the system at higher
frequencies, effectively shaping the noise spectrum which is beneficial in flux
measurements as the noise at higher frequencies is typically more tolerable.

Error diffusion is found to be mean preserving; the mean output signal equals
the mean of the input as the filter coefficients sum to unity. Furthermore, our
analysis shows that the total quantization error comprises two components, one
of which is a filtered version of the other as shown in
Eq.~\ref{eq:quantization-error-z-domain}.

Flux error resulting from quantization and error diffusion is mainly attributed
to the residual spurious correlation between the
quantization error and the scalar concentration $\overline{\varepsilon'\,c'}$.
The flux error is found to be equal to the difference between the values of the
cross-covariance function $R_{ec}(n)$ evaluated at indices 0 and 1, as shown in
Eq.~\ref{eq:error-by-cov-function}. This implies that a larger auto-correlation
of $e(n)$ corresponds to a smaller error, as a one-sample delayed error will
have a stronger correlation with the current $e(n)$. 
The variant of error diffusion proposed in this study diffuses the error per
wind direction, ensuring that the error $e(n)$ remains uncorrelated with the
scalar concentration $c(n)$ even when a non-uniform quantizer is used. This
variant allows for a more flexible choice of quantization parameters, which is
very useful for practical implementations, as will be shown in the
Section~\Ref{sec:implementation}.

\subsection{Performance evaluation of error diffusion}

\begin{figure*}[ht]
\centering
\includegraphics[width=15cm]{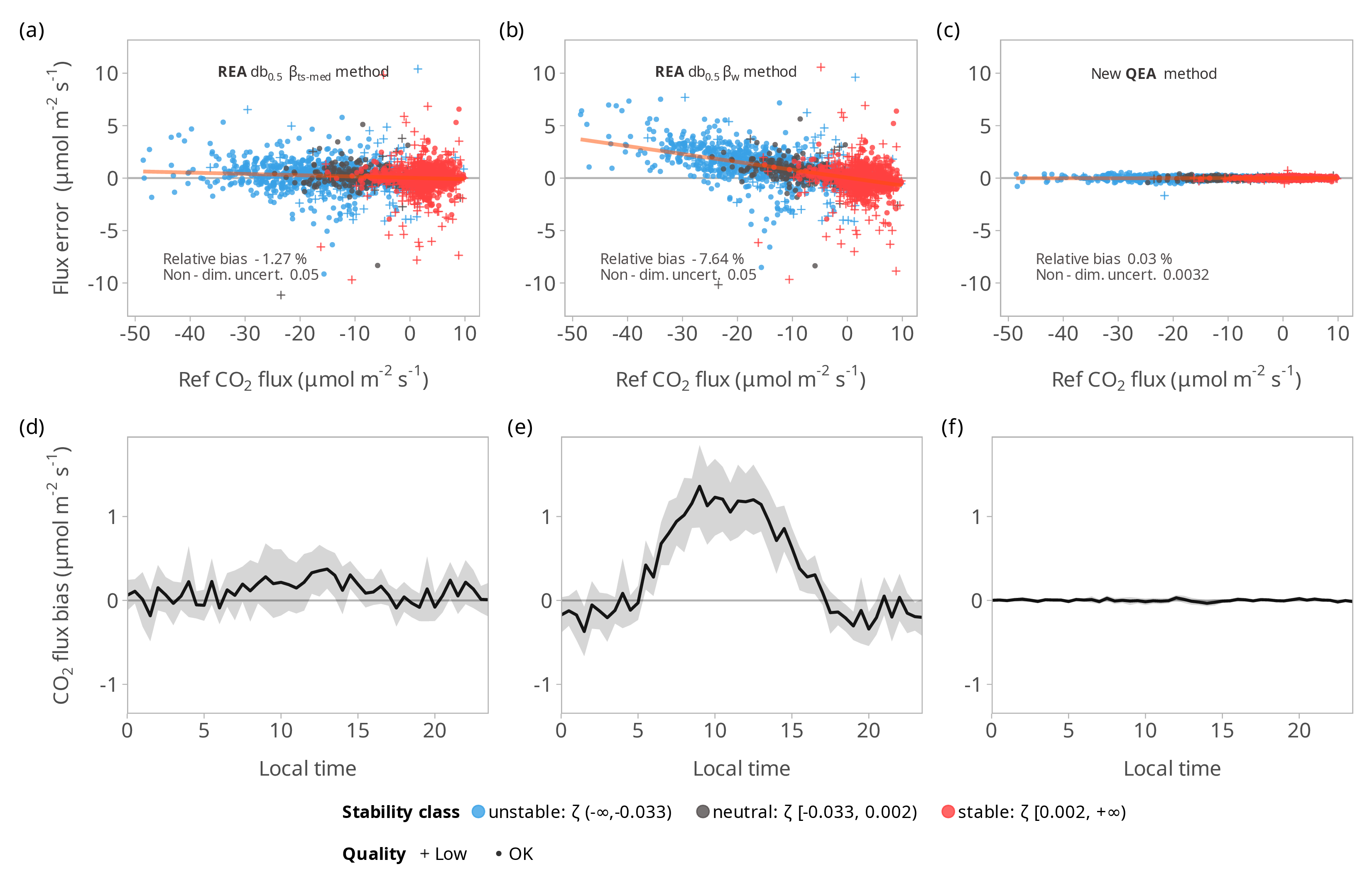}
\caption{Performance evaluation of quantized eddy accumulation. (a)-(c) Scatter
    plots displaying the dependence of flux errors on the reference flux for \chem{CO_2}
    for two established relaxed eddy
    accumulation (REA) variants (\(\beta_{\text{ts-med}}\) and \(\beta_{w}\))
    versus the novel quantized eddy accumulation (QEA) method with error
    diffusion. Flux errors for \chem{CO_2} are calculated as the difference
    relative to eddy covariance reference flux measurements, differentiated by
    atmospheric stability classes: unstable (blue), neutral (gray), and stable
    (red) and two flux quality classes (point shape).  These categories
    illustrate the error's sensitivity to atmospheric stability. Reported
    relative bias is the slope of the error versus the reference value. 
    Reported non-dimensional uncertainty is defined as the standard deviation of
    flux errors normalized by $\sigma_w\,\sigma_c$ for the entire dataset.
    (d)-(f) Diurnal mean of \chem{CO_2} flux errors for each method, with a 95\%
    confidence interval shaded region generated via bootstrapping. This
    demonstrates the temporal variability of biases throughout the day, and
    demonstrates the QEA method's robustness, as reflected in its consistent
    closeness to the reference eddy covariance flux. 
The data shown are for the Braunschweig station. Reported error metrics and
diurnal flux errors are for the quality-filtered dataset.  
For QEA, a quantization threshold $q_t=2.4\sigma_w$  and a full scale value
$w_f=4\sigma_w$
were used.
}
\label{fig:performance-comparison-against-rea}
\end{figure*}

\begin{figure}[ht]
\centering
\includegraphics[width=6.3cm]{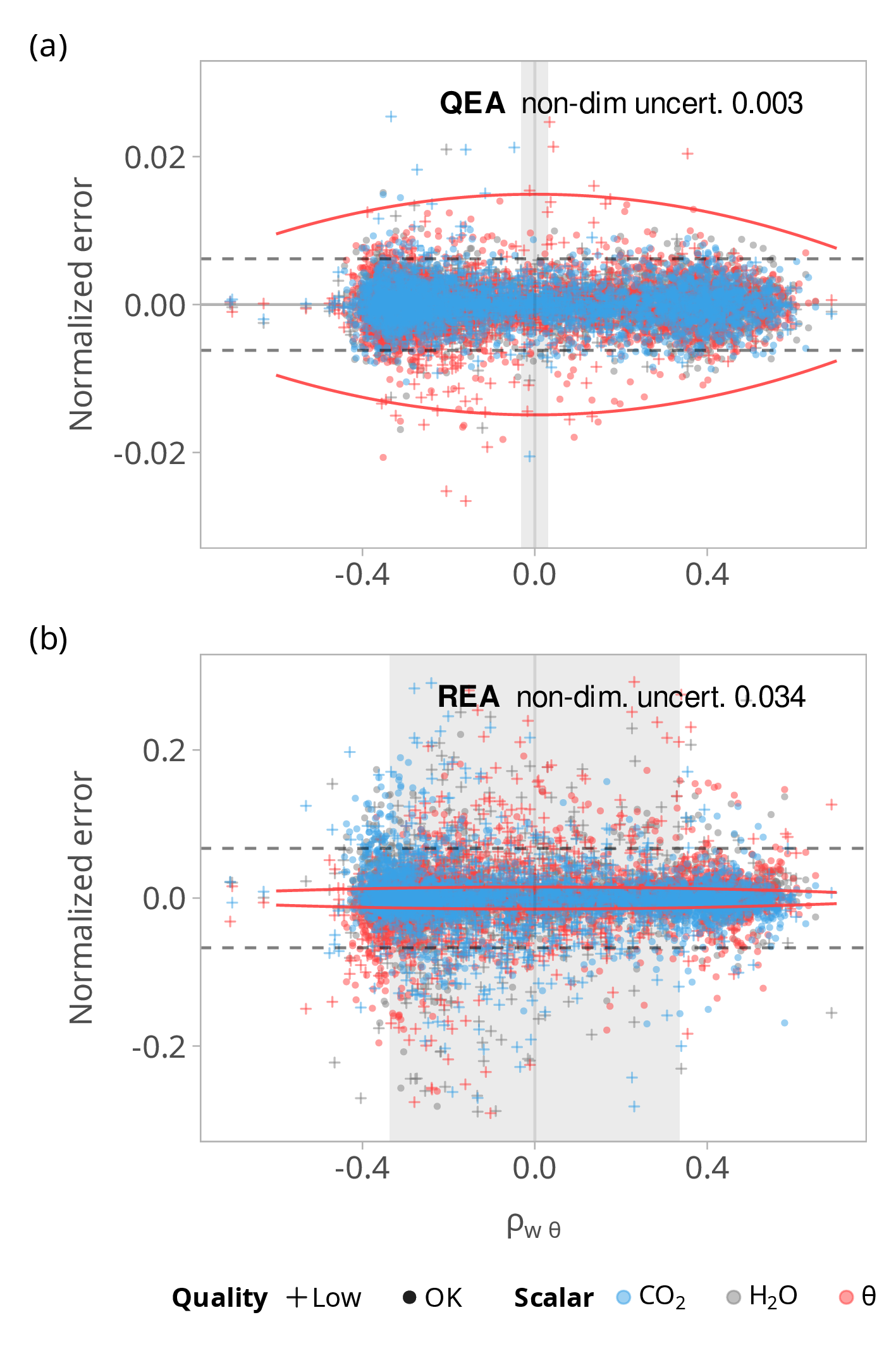}
\caption{
    Normalized flux errors for \chem{CO_2}, \chem{H_2O}, and air temperature
($\theta$) as a function of the correlation coefficient $\rho_{w\,\theta}$.
Panel (a) shows normalized errors of the quantized eddy accumulation (QEA)
method with error diffusion. Panel (b) shows normalized errors of the relaxed
eddy accumulation (REA) method using $\beta_{w}$ with a deadband of
$0.9\sigma_w$.  Dashed horizontal lines indicate $\pm 2~u$, where $u$ is the
non-dimensional uncertainty defined as the standard deviation of the normalized
flux errors for the entire dataset.  The red lines indicate $\pm 2 \times$
theoretical uncertainty expected from the correlation coefficient of two random
samples calculated as $\sigma_r = (1- r^2) / \sqrt{n-3}$.  The shaded areas
represent the range of $\rho_{w\theta}$ where the relative flux uncertainty
exceeds 20\% of the flux value.  }
\label{fig:nondim-uncertainty}
\end{figure}

A simulation-based assessment of the newly developed error diffusion algorithm
indicates a substantial enhancement in accuracy and reduction in uncertainty
over conventional relaxed eddy accumulation methods as shown in 
Fig.~\Ref{fig:performance-comparison-against-rea} and
Fig.~\Ref{fig:nondim-uncertainty}.

We evaluated the performance of the QEA method using high-frequency eddy
covariance data for three scalars (\chem{CO_2}, \chem{H_2O}, and $\theta$),
across two contrasting ecosystems over 18 weeks. We distinguished between
systematic errors, which correlate with the flux values, and random errors,
which do not correlate and are expected to diminish with time averaging.  
To standardize error metrics across different scalars, we normalized the error by
the product of $\sigma_w$ and $\sigma_c$, as shown in
Eq.~\ref{eq:normalized-error}, and defined the non-dimensional uncertainty as
the standard deviation of this normalized error.

The analysis of the simulation results revealed no systematic biases
in QEA fluxes with error diffusion for all tested scalars, sites, and
atmospheric conditions. The slopes of the error against reference flux
values—were consistently below 0.1\%, significantly lower than the average 5\%
bias found with the commonly used REA methods, as shown in
Fig.~\Ref{fig:performance-comparison-against-rea}.

\begin{table*}[ht]
\centering
\caption{Comparative performance evaluation of quantized eddy accumulation (QEA)
    with error diffusion versus conventional relaxed eddy accumulation (REA)
    methods for \chem{CO_2}, \chem{H_2O}, and air temperature ($\theta$) fluxes.
    Performance metrics include error slope (dimensionless), error intercept
    (units of flux), non-dimensional uncertainty, and root mean square error
    (RMSE, units of flux).  Error metrics are calculated as averages for two
    contrasting ecosystems and obtained from a simulation using a high-frequency
    eddy covariance dataset.  For QEA, a quantization threshold
    $q_t=2.4\sigma_w$  and a full scale value $w_f=4\sigma_w$ were used.
}
\label{tab:simulation-results}
\begin{tabular}{llcccr}
\tophline
Method & Scalar & Error slope & Error intercept & Nondim. uncertainty & RMSE \\ 
\middlehline
\multirow{3}{*}{QEA with error diffusion} 
       & \chem{CO_2} & $1.61 \times 10^{-4}$ & $4.19 \times 10^{-4}$  & $2.66 \times 10^{-3}$ & 0.065 \\
       & \chem{H_2O} & $7.73 \times 10^{-5}$ & $1.02 \times 10^{-4}$  & $2.08 \times 10^{-3}$ & $8.90 \times 10^{-3}$ \\
       & $\theta$    & $6.85 \times 10^{-6}$ & $-1.09 \times 10^{-5}$ & $3.21 \times 10^{-3}$ & $3.16 \times 10^{-4}$ \\
 \middlehline
       \multirow{2}{*}{$\mathrm{REA}~\beta_{ts}$}
      & \chem{CO_2} & -0.016 & 0.032                  & 0.430 & 6.891 \\
      & \chem{H_2O} & -0.048 & $-4.19 \times 10^{-3}$ & 0.692 & 1.204 \\
  \middlehline
  \multirow{2}{*}{$\mathrm{REA}~(\mathrm{db}~0.5\sigma)~\beta_{\mathrm{ts-median}}$}
  & \chem{CO_2} & -0.021 & 0.020                 & 0.058 & 1.256 \\
  & \chem{H_2O} & -0.040 & $5.28 \times 10^{-3}$ & 0.056 & 0.169 \\
  \middlehline
  \multirow{3}{*}{$\mathrm{REA}~(\mathrm{db}~0.9\sigma)~\beta_w$}
  & \chem{CO_2} & 0.018 & -0.022                 & 0.056 & 1.165 \\
  & \chem{H_2O} & 0.026 & $-5.99 \times 10^{-3}$ & 0.053 & 0.154 \\
  & $\theta$    & 0.017 & $1.21 \times 10^{-4}$  & 0.060 & $5.10 \times 10^{-3}$ \\
  \middlehline
  \multirow{3}{*}{$\mathrm{REA}~(\mathrm{db}~0.5\sigma)~\beta_w$}
  & \chem{CO_2} & -0.095 & 0.032                  & 0.059 & 1.809 \\
  & \chem{H_2O} & -0.120 & 0.013                  & 0.056 & 0.243 \\
  & $\theta$    & -0.075 & $-1.64 \times 10^{-4}$ & 0.062 & $8.54 \times 10^{-3}$ \\
\bottomhline
\end{tabular}
\belowtable{} %
\end{table*}

\begin{table*}[ht]
\centering
\caption{Average improvement ratio of quantized eddy accumulation (QEA) method
over relaxed eddy accumulation (REA) variants. The table shows the mean
enhancement of QEA over REA variants for each error metric.  The improvement
ratio for each error metric is calculated by dividing the average metric value
for REA across three scalars and two stations by the corresponding average value
for QEA. Ratios larger than one indicate areas where QEA shows significant
advantages }
\label{tab:simulation-results-rel}
\begin{tabular}{lrrrr}
  \hline
Method & RMSE & Error intercept & Error slope & Nondim. uncertainty \\ 
  \hline
  $\mathrm{REA}~(\mathrm{db}~0.9\sigma)~\beta_{\mathrm{ts}}$ & 264 & 1070 & 133 & 499 \\ 
  $\mathrm{REA}~(\mathrm{db}~0.9\sigma)~\beta_{\mathrm{ts-median}}$ & 25 & 185 & 334 & 23 \\ 
  $\mathrm{REA}~\beta_{ts}$ & 173 & 200 & 546 & 219 \\ 
  $\mathrm{REA}~(\mathrm{db}~0.9\sigma)~\beta_w$ & 21 & 68 & 401 & 21 \\ 
  $\mathrm{REA}~(\mathrm{db}~0.5\sigma)~\beta_{\mathrm{ts-median}}$ & 28 & 67 & 537 & 23 \\ 
  $\mathrm{REA}~(\mathrm{db}~0.5\sigma)~\beta_{ts}$ & 947 & 1788 & 1290 & 618 \\ 
  $\mathrm{REA}~(\mathrm{db}~0.5\sigma)~\beta_w$ & 41 & 108 & 2161 & 26 \\ 
   \hline
\end{tabular}
\end{table*}

The QEA method have consistently demonstrated minimal random errors, with a
non-dimensional uncertainty below 0.004 for all scalars, significantly lower
than the 0.05 observed with the best-performing REA variant as shown in
Fig.~\Ref{fig:nondim-uncertainty}.  The uncertainly of QEA is smaller than the
theoretical uncertainty expected from the correlation coefficient of two random
samples, calculated as $\sigma_r = (1- r^2) / \sqrt{n-3}$
\citep{gnambsBriefNoteStandard2022,bonettSampleSizeRequirements2000} and shown
as red lines on Fig.~\Ref{fig:nondim-uncertainty}. 
This can be attributed to the auto-correlation of the
quantizer error that makes the difference $R_{ec}(0) - R_{ec}(1)$ very small.

Importantly, the accuracy of QEA was not influenced by the time of day,
atmospheric stability, or flux quality metrics, unlike the REA methods, as shown
in  Fig.~\Ref{fig:performance-comparison-against-rea}.  These findings, along
with several other error metrics, are summarized in
Table~\ref{tab:simulation-results} and Table~\ref{tab:simulation-results-rel},
demonstrating that QEA outperforms REA across all examined conditions and
metrics. The relative improvement ratio of QEA over REA variants reported in
Table~\ref{tab:simulation-results-rel} shows wide variation across the different
REA methods and scalars. However, QEA has a consistent improvement over all REA
methods and scalars with the error slope showing over 100-fold improvement
compared to the best-performing REA variant and up to 1000-fold improvement for
some variants and scalars. Additional figures are provided in the
supplementary material showing more detailed comparisons of the performance of
QEA and REA methods across different scalars and stations.

\subsection{Optimal quantization parameters}
QEA with error diffusion has two parameters: the quantization threshold $q_{t}$
and the quantized full-scale value $w_{f}$.  The choice of these parameters
allows for minimizing the flux error, maximizing the concentration difference
$\Delta C$ in accumulation reservoirs, reducing the error due to nonzero mean
vertical wind velocity, and controlling the switching rates of the sampling
valves.

\begin{figure}[ht]
\centering
\includegraphics[width=8.3cm]{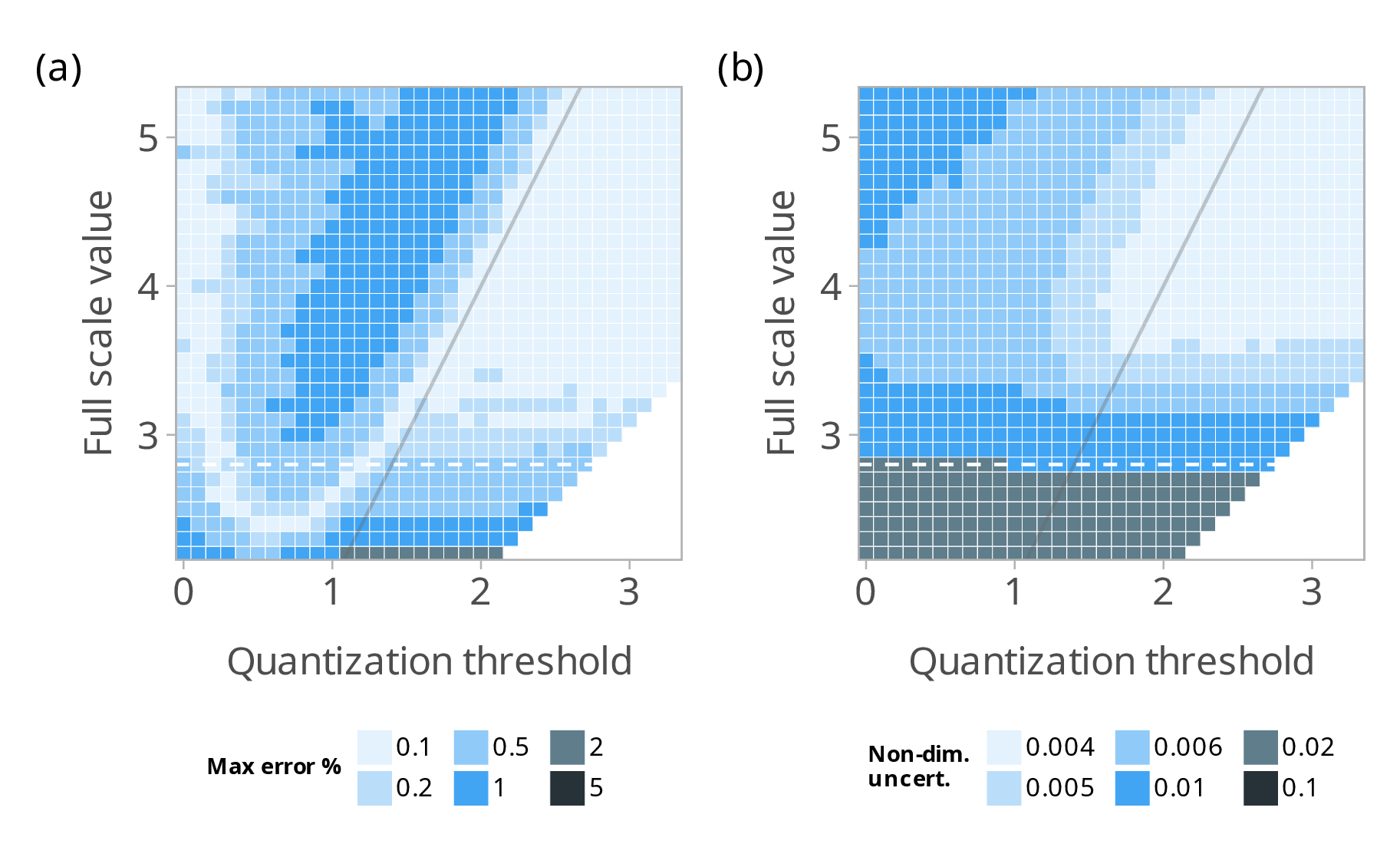}
\caption{
Evaluation of parameter space for systematic bias and uncertainty for
\chem{CO_2} flux measurements with QEA.  Panel (a) shows a heatmap that
illustrates the systematic bias, represented by the slope of the linear fit, as
a function of the quantization threshold and full-scale value. The color
intensity corresponds to the maximum magnitude of the systematic bias in each
class. 
A white dashed line at a full-scale value of $2.7$ and a solid line with a slope
of $2$ indicate the boundaries where systematic biases exceeds 0.2\%.  
Panel (b) presents a heatmap of the maximum
non-dimensional uncertainty which indicates error variability.  
These metrics
are calculated based on simulations utilizing the entire dataset for the
Braunschweig station.
}
\label{fig:parameter-space-heatmap}
\end{figure}

The main objective of the evaluation presented here was to establish optimal
ranges for the quantizer parameters that optimize for the above goals. These
goals were assessed through numerical simulations for three different scalars
over a broad range of parameters.
Simulation results indicated that, for all scalars, the flux error remained
consistently low across a wide range of $q_{t}$ and $w_{f}$ combinations, as
shown in Fig.~\Ref{fig:parameter-space-heatmap}. Generally, the error stayed
below 1\% for almost any combination of parameters, provided that the full-scale
value was greater than $2.7 \sigma_w$. Notably, a large number of parameter
combinations resulted in errors smaller than 0.1\%. The random errors showed
similar trends, with the non-dimensional uncertainty typically being lower for
parameter combinations that minimized systematic errors.  The error diffusion
algorithm variant used in this simulation diffuses quantization errors per wind
direction. This variant was found to be more effective in reducing errors and
offers greater flexibility in the choice of quantization parameters, as will be
shown in Section~\ref{sec:implementation}.

\begin{figure}[ht]
\centering
\includegraphics[width=6.3cm]{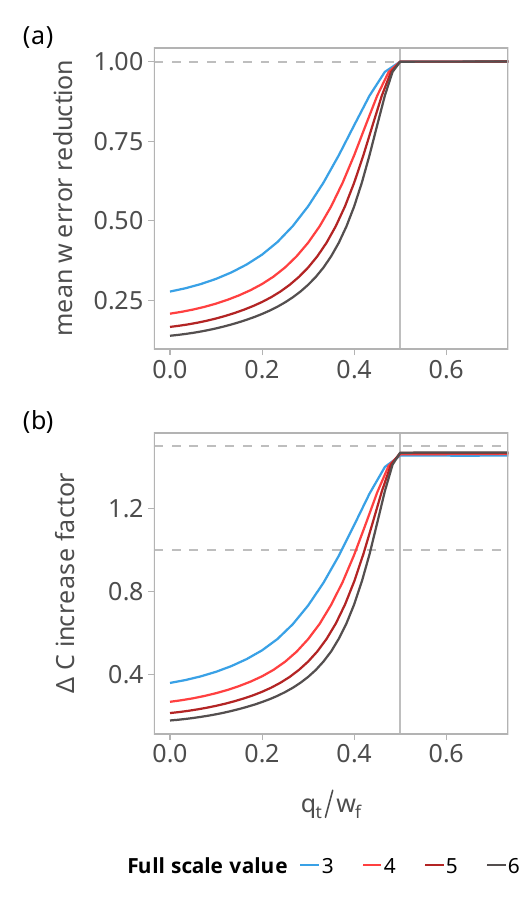}
\caption{
Nonzero mean wind error reduction and $\Delta C$ increase of error diffusion.
(a) Nonzero mean $w$ error reduction factor against scaled quantization
threshold, for full-scale values 3-6, illustrating the impact of quantization on
errors associated with nonzero $\overline{w}$. The reduction factor is
calculated as the ratio of the mean of absolute wind velocity $\overline{|w|}$
to the mean of absolute quantized velocity $\overline{|w_q|}$. Values below one
indicate a reduction in the error. (b) \(\Delta C\) increase factor, calculated
as the slope of a linear fit of $\Delta C$ with the given quantization threshold
to $\Delta C$ with a quantization threshold of zero. Dashed horizontal lines
shown at y=1 and y=1.5. $q_t/w_f$ is the scaled quantization threshold which we
defined as the ratio of the quantization threshold to the full-scale value.  }
\label{fig:w-mean-error-reduction}
\end{figure}

Adjusting the quantization threshold can improve the signal-to-noise ratio by
increasing the difference in concentration  $\Delta C$ between the two
accumulation reservoirs, as illustrated in
Fig.~\Ref{fig:w-mean-error-reduction}.  The key driving factor in increasing
$\Delta C$ is the ratio between the quantization threshold and the quantized
full-scale value, referred to as the scaled quantization threshold.  Values
greater than 0.5 can increase $\Delta C$ by up to 150\%, akin to introducing a
deadband in REA measurements, which leads to the selective accumulation of
eddies with concentrations further away from the mean.  However, unlike REA
where increasing the deadband is accompanied by an increase in flux uncertainty
\citep{oncleyVerificationFluxMeasurement1993,patteyAccuracyRelaxedEddyaccumulation1993},
selecting a higher quantization threshold to increase $\Delta C$ does not reduce
the accuracy of fluxes measured with QEA as seen from
Fig.~\ref{fig:parameter-space-heatmap}.

Reducing the scaled quantization threshold can help minimize errors related to
non-zero mean vertical wind velocity.  This reduction coincides with a decrease
in $\Delta C$, as shown in Fig.~\Ref{fig:w-mean-error-reduction}. The reduction
of errors caused by non-zero mean wind velocity is driven by increased wind
variance (and thus, $\overline{|w|}$), which is inversely correlated with flux
error magnitude under conditions when the mean wind is not zero
\citep{emadTrueEddyAccumulation2023}.
A scaled quantization threshold below 0.5 is effective in decreasing these
errors as shown in Fig.~\Ref{fig:w-mean-error-reduction}. A threshold set to
zero can remove the influence of non-zero mean vertical wind velocity entirely,
which is a unique advantage of QEA over TEA.

\begin{figure}[ht]
\centering
    \includegraphics[width=8.3cm]{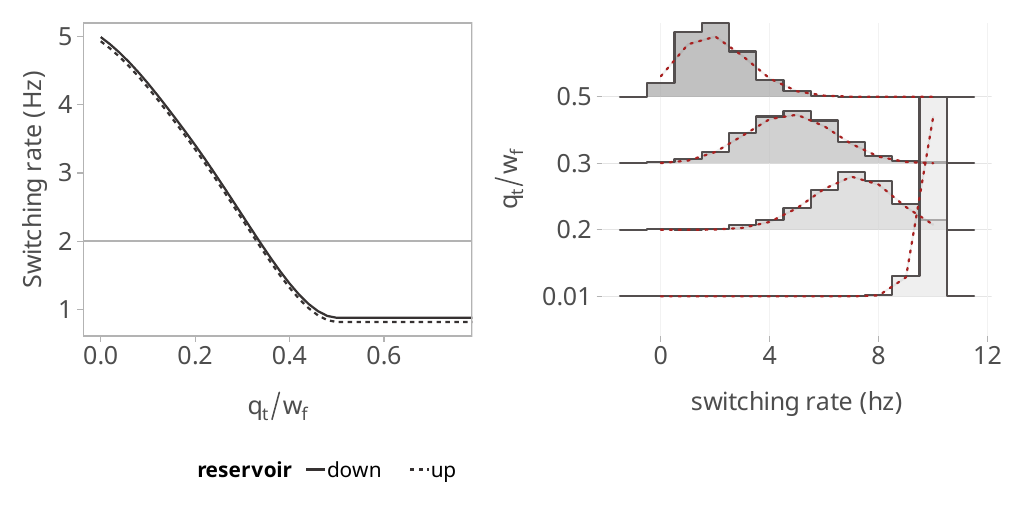}
\caption{
Valve switching dynamics. The left panel illustrates the relationship between
the scaled quantization threshold and the average valve switching rate (Hz),
distinguishing between reservoirs ('down', solid line) and ('up', dashed line).
The right panel presents histograms of the distribution of valve switch rates
(as events per second) for four distinct scaled quantization thresholds
simulated from a 30-minute high-frequency wind measurements dataset.  
Dashed lines indicate the theoretical binomial distribution for the valve
switching events per second, with the success probability corresponding to the
average valve switching rate and the size parameter as the number of samples
taken per second.  }
\label{fig:valve-open-distribution}
\end{figure}

Additionally, adjusting the scaled quantization threshold affects how often
sampling valves switch and the accumulated sample volumes.  Higher thresholds
lead to less frequent switching; for example, with a threshold of 0.5, the
updraft and downdraft valves switch on average once per second, as shown in
Fig.~\Ref{fig:valve-open-distribution}.

\subsection{Implementation of quantized eddy accumulation with error diffusion}
\label{sec:implementation}

Quantized eddy accumulation with error diffusion can be implemented using a
varying number of quantization levels and based on different conditioning
variables. However, the simplest and most useful implementation is achieved
using two reservoirs that accumulate air at a constant flow rate, similar to the
requirements of the relaxed eddy accumulation method. 
Since each quantization level is accumulated in a separate reservoir, no
proportional control of the flow is required.
We will discuss the specifics of such an implementation here.

The error diffusion algorithm we developed for this implementation is described in
Algorithm \ref{algo:qea-with-error-diffusion}. This implementation requires only
a constant airflow rate for air sampling and the channelling of this airflow
into either updraft or downdraft reservoirs depending on the sign of quantized
wind velocity. 
In this algorithm, the quantization error is incorporated in a feedback loop
with the measured vertical wind velocity before it is quantized into three
discrete levels corresponding to updraft, downdraft, and no sampling. The
quantization error is managed separately for updraft and downdraft. This
separation has been found to improve the stability of the flux errors and allows
for a wider range of quantization parameters that produce accurate flux
estimates.

\begin{algorithm}
\caption{Quantized eddy accumulation with error diffusion}
\label{algo:qea-with-error-diffusion}
\begin{algorithmic}[1]
\REQUIRE vertical wind speed array $w$, quantization threshold $q\_threshold$, and wind
quantized full-scale value $w\_full$

\STATE $N \leftarrow \text{length}(w)$
\STATE $residual\_error\_up \leftarrow 0$
\STATE $residual\_error\_down \leftarrow 0$
\FOR{$i = 1$ \TO $N$}
    \IF{$w[i] > 0$}
    \STATE $w_{mod} \leftarrow w - residual\_error\_up$ 
    \STATE $w_q[i] \leftarrow Q(w_{mod},~q\_threshold,~w\_full)$
    \STATE $residual\_error\_up \leftarrow w_q[i] - w_{mod}$
    \ELSE
    \STATE $w_{mod} \leftarrow w - residual\_error\_down$ 
    \STATE $w_q[i] \leftarrow Q(w_{mod},~q\_threshold,~w\_full)$
    \STATE $residual\_error\_down \leftarrow w_q[i] - w_{mod}$
    \ENDIF
\ENDFOR
\RETURN $w_q$
\end{algorithmic}
\end{algorithm}

The quantized $w$ can take one of three values: $-w_f$, 0, or $w_f$,
where $w_f$  is the quantized full-scale level.
The accumulated volume in each reservoir at the end of the
averaging interval is calculated as $V = A\, n_q \, T_s$, where $A$ represents a
flow scaling factor (\unit{m^3\,s^{-1}}), $n_q$ is the number of sampling events
in the averaging interval, and $T_s$ is the duration of sampling for each event
(\unit{seconds}).

The choice of the quantization threshold ($q_t$) and the quantizer's
full-scale setting ($w_f$) plays a crucial role in error control,
signal-to-noise ratio enhancement, and reduction of residual mean vertical wind
velocity errors.
Setting $q_t$ to zero allows for an unbiased estimation of $\bar{c}$, as it
leads to $\overline{c^\uparrow w_q^\uparrow} =
\overline{c^\uparrow}~\overline{w_q^\uparrow}$. 
This choice effectively eliminates errors associated with non-vanishing mean
wind velocity, which provides an advantage over traditional eddy accumulation
methods.

Referring to Fig.~\ref{fig:parameter-space-heatmap}, we can identify the
optimal quantization parameters $q_t$ and $w_f$. In general,
to minimize the errors in fluxes calculated via QEA, it is recommended to use
a quantized full-scale value in the range $[4,5]~\sigma_w$ and a quantization
threshold in the range 0.55 to 0.85 of the full-scale value.
For instance,
setting $q_t$ to $2.5 \sigma_w$ and the full-scale value to $4 \sigma_w$
predicts an average error under $0.1\%$. 
Since $q_t$ and $w_f$ depend on the unknown $\sigma_w$ of the current averaging
interval; an estimate from the preceding interval can be used.
The heatmap of Fig.~\ref{fig:parameter-space-heatmap} illustrates that error
diffusion is robust to slight variations in $\sigma_w$ which supports the use of
past data to define the parameters of the
current averaging interval. 
The chosen example parameters suggest nearly a
1.4-fold increase in $\Delta C$ and an average valve switching rate of about 2
Hz as seen from figures \ref{fig:w-mean-error-reduction} and
\ref{fig:valve-open-distribution}.

The quantized vertical wind velocity, $\overline{w_q}$, is expected to equal the
vertical wind velocity, $\overline{w}$, because quantization with error
diffusion preserves the mean, as demonstrated earlier.
The error associated with nonzero $\overline{w_q}$ in QEA can be addressed in
three ways:
i) Similar to true eddy accumulation methods, an estimate of $\overline{c}$ can
be derived from available quantities by averaging $c^\uparrow$ and
$c^\downarrow$ with proper weights as shown in
Eq.~\ref{eq:flux-calculation-total}.
ii) QEA offers an advantage over conventional EA methods. By setting $q_t
= 0$, we can calculate $\overline{c}$ from measurements. This, however, may
reduce $\Delta C$, as shown in Fig. \ref{fig:w-mean-error-reduction}. Still, it
can be justified if the analytical instrument is sufficiently accurate.
iii) Under stationary conditions, the error associated with a non-zero mean
vertical wind velocity (\(\bar{w}\)) is limited to the ratio
\(\bar{w}/\overline{|w|}\) \citep{emadTrueEddyAccumulation2023}. Therefore,
error diffusion presents a novel opportunity to minimize this error by
amplifying the wind's variance, which subsequently raises the mean absolute wind
velocity (\(\overline{|w|}\)). This strategy effectively mitigates the error
related to a non-zero $\overline{w}$, as demonstrated in Figure
\ref{fig:w-mean-error-reduction}.

In a typical implementation of the QEA method, real-time vertical wind velocity is
acquired and then adjusted through an online planar fit to align the
coordinates to the streamline coordinates, as described in \citep{siebickeTrueEddyAccumulation2019}. 
We recommend subtracting the mean wind of the previous interval (differencing)
to minimize residual $\bar{w}$ and achieve a more symmetric distribution of
$\bar{w}$ around zero. Subsequently, the wind speed is modified
using the previous quantization error and then quantized 
using Eq.~\Ref{eq:quantizer-function} according to
Algorithm~\ref{algo:qea-with-error-diffusion}.
If the quantized $w$ is non-zero, the flow is directed into the corresponding
reservoir.  
It is important to emphasize that the requirement for flow consistency applies
only within each averaging interval and reservoir. Therefore, variations in flow
are permissible between updraft and downdraft and across different averaging
intervals. 

The implementation of the new QEA method with error diffusion presents several
critical technical challenges. A key challenge is the need for fast-switching
valves, which are essential for directing the flow between the two accumulation
volumes. 
While we have demonstrated that the mean switching rate can be reduced to as
low as 2 Hz, the switching must occur within milliseconds to ensure accuracy, as
any time lag cannot be corrected in post-processing. The flow perturbations
caused by the switching mechanism must also be minimized. Additionally, the
system must be designed to minimize dead volumes and time lags. The accumulation
volumes need to be optimized to provide sufficient sample quantities for the
analyzer while accommodating the dynamic range of the accumulated samples.
Many of these challenges are shared with the
relaxed eddy accumulation method and are well documented in the literature
\citep{ammannApplicabilityRelaxedEddy1998}.
Meeting these requirements is fundamental to the successful deployment and
operation of the QEA method. 

By the end of the averaging interval, the flux can be calculated using
Eq.~\Ref{eq:flux-calculation-total} and corrected for nonzero $\overline{w_q}$ if
needed using Eq.~\Ref{eq:base-flux-f1}.
Since the measured quantities here are means rather
than fluctuations, eddy accumulation methods are more robust to
frequency losses due to the analyzer response and typically will not require
spectral corrections \citep{emadOptimalFrequencyResponseCorrections2023}.

\conclusions  %

In this paper, we introduced the quantized eddy accumulation (QEA) method with
error diffusion, a new direct micrometeorological method with minimal
implementation requirements. We framed the problem of flux measurement with
conditional sampling at a constant flow rate as measuring the flux using a
quantized wind signal. The conventional relaxed eddy accumulation (REA) method
was identified as a special case of this framework, with biases in REA linked to
the unaccounted flux portion transported by the covariance between the
quantization error and scalar concentration.

This new formulation enabled us to develop an error diffusion algorithm that
feeds quantization errors back into the signal, thereby driving the unaccounted
flux term due to quantization errors to zero and eliminating the need for the
empirical coefficient $\beta$.  QEA with error diffusion aligns with eddy
covariance and true eddy accumulation as a direct method while offering the
distinct advantage of enhancing the signal-to-noise ratio without compromising
accuracy.

Our analysis and numerical simulations using high-frequency EC data from two
contrasting ecosystems demonstrated that QEA has achieved unbiased flux estimates
with errors below 0.1\% over a wide range of
quantization parameters and atmospheric conditions. 
Key technical challenges to the successful implementation of QEA include the
need for fast-switching valves to direct flow between accumulation volumes,
minimizing flow perturbations, and reducing time lags within the system.

The new method provides simple and reliable means for accurate flux measurements
of challenging atmospheric constituents and has the potential to advance our
understanding of atmospheric chemistry and earth science.

\dataavailability{
The datasets generated and analyzed in this study, along with the full code
implementation, are available at \cite{emadDataSupportCode2024}
and the github repository 
\url{https://github.com/anasem/quantized-eddy-accumulation}

Raw data for Braunschweig station that were
used as input for the simulation are available at
\cite{emadHighfrequencyWindTs2024}.
} %

\begin{acknowledgements}
I am grateful to the editor and reviewers for their comments and
suggestions, which significantly improved this manuscript.

I acknowledge with gratitude the support of the Bioclimatology group led by Alexander Knohl at the
University of Göttingen. I thank Lukas Siebicke for his guidance and valuable
discussions on conditional sampling methods, and Anne Klosterhalfen for her
assistance with the Hainich data.

I acknowledge the support by the German Federal Ministry of Education and
Research (BMBF) as part of the European Integrated Carbon Observation System
(ICOS).
This work was partially funded by the Leibniz Association (Leibniz Collaborative
Excellence Project ISO-SCALE), the Ministry of Lower Saxony for Science and
Culture (MWK), and the Deutsche Forschungsgemeinschaft (INST 186/1118-1 FUGG).

\end{acknowledgements}

\bibliographystyle{unsrtnat}
\bibliography{references}  

\end{document}